\documentclass[aps,prl,superscriptaddress,showpacs,longbibliography,floatfix,nofootinbib,twocolumn]{revtex4-1}

\usepackage[utf8]{inputenc}
\usepackage{slashed}
\usepackage{color}
\usepackage{graphicx}   % Include figure filed
\usepackage{bm}
\usepackage{amsmath}
\usepackage{amsfonts}
\usepackage{hyperref}
\def\be{\begin{equation}}
\def\ee{\end{equation}}

\begin{document}

\title{Vanishing Polyakov Loop for QCD with Twelve Massless Quarks}

\author{Seth Grable}
\affiliation{Department of Physics, University of Colorado, Boulder, Colorado 80309, USA}
\author{Paul Romatschke}
\affiliation{Department of Physics, University of Colorado, Boulder, Colorado 80309, USA}
\affiliation{Center for Theory of Quantum Matter, University of Colorado, Boulder, Colorado 80309, USA}
%\emailAdd{paul.romatschke@colorado.edu}

\begin{abstract}
  We consider continuum-formulation QCD in four dimensions with twelve massless fundamental quark flavors. Splitting the SU($N$) gauge field into background and fluctuation parts, we use well-developed techniques to calculate the one-loop effective action for the theory. We find that for constant self-dual background field-strength tensor the notorious infrared divergences of the effective action cancel between gauge and matter sectors if the number of massless quark flavors is exactly $N_f=4 N$. The ultraviolet divergencies of the effective action are non-perturbatively renormalized with a $\beta$-function that matches the known perturbative result in the high energy limit. The resulting UV- and IR-finite effective action possesses a non-trivial saddle which has lower free energy than the perturbative vacuum, and for which the expectation value of the Polyakov loop vanishes. Inclusion of finite temperature effects points to the presence of a first-order phase transition to the perturbative vacuum with a calculable critical temperature. 
\end{abstract}

\maketitle

\section{Introduction}

Quantum Chromodynamics (QCD) is notorious for being a quantum field theory that is difficult to solve on large length scales, and complete analytic solutions are only known in lower dimensions without fermions, see e.g. Ref.~\cite{Gross:1980he}. Unlike at short length-scales, where perturbative approaches have proven tremendously successful (see e.g. \cite{Gross:1973id,Politzer:1973fx,Kapusta:1979fh,vanRitbergen:1997va,Brambilla:1999xf,Bauer:2000yr,Kajantie:2002wa,Moch:2004pa,Laine:2006ns,Kurkela:2009gj,Gorda:2018gpy,Fernandez:2021jfr}), knowledge about the infrared behavior of QCD comes mostly from numerical approaches, mainly lattice QCD \cite{Wilson:1974sk,deForcrand:2002hgr,HPQCD:2003rsu,Kaczmarek:2005ui,Aoki:2006we,BMW:2008jgk,NPLQCD:2012mex,Borsanyi:2013bia,HotQCD:2014kol,deForcrand:2014tha,Borsanyi:2020mff}, as well as models for individual aspects of QCD that are too numerous to mention.

Widely explored and successful approaches to the non-perturbative structure of QCD broadly fall into two categories: studies of supersymmetric (SUSY) QCD \cite{Intriligator:1995au} and QCD with extra heavy matter \cite{Unsal:2007jx,Poppitz:2009uq,Cherman:2016hcd}. The SUSY case involves extra (colored) scalar fields which together with known SUSY dualities can be used to set up a derivation of confinement that is within a weak-coupling regime. In the case of extra heavy matter, one adds at least one additional heavy adjoint fermion to QCD, and then compactifies the theory on a circle with appropriate boundary conditions, leading to a confining theory through a controlled semi-classical expansion.

By contrast, the motivation for our work is provided by the search of a (potentially approximate) solution to continuum QCD that captures the main qualitative features such as confinement in the IR and asymptotic freedom in the UV, without adding extra assumptions or other ingredients\footnote{As will become clear in the main text, in this work we will need to ``bend'' our goal by adding additional fundamental fermion species beyond those present in physical real-world QCD.}

A priori, it is not clear that such a solution exists at all. However, if such a solution does exist, it cannot be based on a perturbative weak-coupling expansion, because QCD becomes strongly coupled towards long length scales. So the putative solution must arise from non-perturbative methods.

Curiously, tremendous progress on obtaining non-perturbative solutions to QCD has been made using semi-classical approximations, cf. ~\cite{Savvidy:1977as,Nielsen:1978rm,Yildiz:1979vv,Leutwyler:1980ma}, see also \cite{Voskresensky:1992rm}. Using methods going back to Euler and Heisenberg in the 1930s (see Ref.~\cite{Dunne:2012vv} for an excellent review), it was found that -- unlike QED -- pure Yang-Mills contained infrared instabilities in the case of a constant background chromomagnetic field \cite{Savvidy:1977as,Nielsen:1978rm}, but these instabilities were rendered more benign for self-dual field configurations \cite{Leutwyler:1980ma}. However, even for constant self-dual background field, massless excitations that Leutwyler dubbed 'chromons' remain for pure Yang-Mills \cite{Leutwyler:1980ma}, such that the resulting effective potential still contains infrared divergencies that are sometimes regulated by hand \cite{Savvidy:2022jcr}.

In our opinion, the approach pioneered by Refs.~\cite{Savvidy:1977as,Nielsen:1978rm,Leutwyler:1980ma} is extremely promising, especially if incorporating resummation schemes such as R2 outlined in Refs.~\cite{Romatschke:2019rjk,Romatschke:2019ybu} can be shown to lead to a infrared-finite effective action without any need for regulators.

Here, instead of performing this exciting but difficult task, we use a trick to remove the remaining infrared divergence in the effective action encountered by \cite{Leutwyler:1980ma}. Using the same techniques to include massless Dirac fermions in the fundamental representation, we explicitly show that the fermions also contribute an infrared divergence to the effective action, but with an opposite sign than the gluons, cf. \cite{Kay:1983an}. Thus, for a certain number of massless fermion species only, we are able to cancel the infrared divergence between fermionic and gluonic sectors, and obtain an effective action free of infrared divergencies. Note that the required number of fermions for this to happen does \textit{not} correspond to supersymmetry, nor any other symmetry that we are aware of\footnote{It should be pointed out that the cancellation of UV divergencies in the vacuum energy between equal numbers of fermionic and bosonic degrees of freedom was noted by W.~Pauli much before the properties of supersymmetry were discovered, see e.g. the discussion in Ref.~\cite{Kamenshchik:2018ttr}. We suspect that our observed cancellation of IR divergencies could be lead to a similar development.} . 

With the infrared divergences canceled, we can renormalize the UV divergencies of the theory and obtain a finite effective action, and with this effective action, we are able to analytically calculate observables in the infrared limit of the theory.

Before delving into the details of the calculation, we would like to stress that because of the somewhat accidental cancellation of infrared divergencies between the fermionic and gluonic sectors of the theory, our predictive power over the theory is limited. This is because any deformation that changes the masses, such as explicit fermion mass terms, finite temperature, or finite density, ruins the delicate cancellation of infrared divergences that our construction relies on. In our opinion, this is just a reflection of the fact that we do not \textit{really} cure the infrared divergences, we just make them cancel. Nevertheless, we believe that for those situations where  the infrared cancellation occurs, results are sufficiently interesting to discuss our findings.

\section{Calculation}

We consider an SU(N) gauge theory coupled to $N_f$ massless Dirac fermions in $d=4-2\varepsilon$ Euclidean dimensions with partition function
\be
Z=\int {\cal D}A {\cal D}\bar\psi {\cal D}\psi e^{-S_E}\,,
\ee
where the Euclidean action is given by
\be
S_E=\int dx \left[\frac{1}{4 g_0^2} F_{\mu\nu}^a F_{\mu\nu}^a+\bar\psi_f \slashed{D} \psi_f\right]\,,
\ee
and where $g_0$ is the bare Yang-Mills coupling constant.
Here
\be
\label{fmunu}
F_{\mu\nu}^a=\partial_\mu A_\nu^a-\partial_\nu A_\mu^a+f^{abc}A_\mu^b A_\nu^c
\ee
is the Yang-Mills field strength tensor with $A_\mu^a$ the gauge potential and $\psi_f=\left(\psi_1,\psi_2,\ldots,\psi_{N_f}\right)$ are the $N_f$ Dirac fermions in the fundamental representation of SU(N). $\slashed{D}=\gamma_\mu \left(\partial_\mu-i A_\mu^a t^a\right)$ is the Dirac operator with the Euclidean gamma matrices $\gamma_\mu$ fulfilling the commutation relation $\left\{\gamma_\mu,\gamma_\nu\right\}=2 \delta_{\mu\nu}$. The SU(N) generators $t^a$ and structure constants $f^{abc}$ are taken to fulfill ${\rm tr}\, t^{a}t^{b}=\frac{\delta^{ab}}{2}$ and $\left[t^a,t^b\right]=i f^{abc}t^c$.

Our basic assumption is that the partition function is dominated by a non-trivial non-perturbative classical saddle point configuration of the gauge field. However, owing to the absence of a small dimensionless expansion parameter, the semi-classical expansion can not be systematically improved, and hence is uncontrolled. Splitting the gauge field into a background field $\bar A_\mu^a$ and fluctuations around it as
\be
\label{split}
A_\mu^a(x)=\bar A_\mu^a(x)+a_\mu^a(x)\,,
\ee
we note that the background field is a fixed, but so far arbitrary function of space-time, so that the partition function takes the form
\be
Z=\int d\bar A {\cal D}a {\cal D}\bar\psi {\cal D}\psi e^{-S_E}\,.
\ee
Expanding the action as $S_E=S_0+S_1+S_2+\ldots$ with the index denoting powers of the fluctuation fields, such that e.g.
\be
S_2=\frac{1}{4 g_0^2}\int dx\left[\left(d_\mu^{ac}a_\nu^c-d_\nu^{ac}a_\mu^c\right)^2+2 \bar F_{\mu\nu}^a f^{abc} a_\mu^b a_\nu^c\right]\,,
\ee
 where $d_\mu^{ac}=\delta^{ac}\partial_\mu +f^{abc} \bar A_\mu^b$ is the background gauge-covariant derivative and $\bar F_{\mu\nu}^a$ is (\ref{fmunu}) for $A_\mu^a=\bar A_\mu^a$. We thus obtain
\be
Z=\int d\bar A e^{-S_{\rm eff}[\bar A]}\,,\quad S_{\rm eff}[\bar A]=S_0[\bar A]+S_2[\bar A]
\ee
where $S_{2}[\bar A]$ is the effective potential resulting from integrating out all fluctuations resulting from the action $S_E-S_0=S_1+S_2+\ldots$. As it stands, the path integral over the fluctuating gauge field $a_\mu^a$ is ill-defined, because $S_E$ is invariant under gauge-transformations, implying flat integration directions along which $S_E$ is constant. The accepted remedy for this situation is to gauge-fix the action. We employ Weinberg's gauge fixing condition \cite{weinberg1995quantum}, which implies adding a gauge-fixing and ghost term of the form (truncated at second order)
\be
S_{\rm gf}=\int dx \frac{\left(d_\mu^{ac} a_\mu^c\right)^2}{2g_0^2}\,,\quad 
S_{\rm ghost}=\int dx d_\mu^{ac} \bar c^{c} d_\mu^{ab}c^b\,,
\ee
to the action, where $\bar c,c$ are the ghost fields.

Unfortunately, an exact calculation of the effective action is beyond our capabilities, so we have to resort to approximations. Our first approximation is to truncate the expansion of $S_E$ at second order in the fluctuation fields, by which we mean $a_\mu^a,\bar\psi,\psi,\bar c,c$. This corresponds to the one-loop approximation of the effective action, or R0-level approximation for O(N) models \cite{Romatschke:2019ybu}. It is important to recall that the one-loop effective action corresponds to the leading order semi-classical expansion of the theory, which resums an infinite number of weak-coupling contributions, cf. \cite{Moshe:2003xn}. Therefore, despite not being an exact solution, this one-loop approximation is nevertheless non-perturbative. However, as explained above, the absence of a small parameter makes the semi-classical expansion uncontrolled.

Since we are looking for saddle point solutions for the background gauge field $\bar A_\mu$, this implies that $S_1=0$, so that the one-loop effective action can be calculated in closed form.

One finds
\be
S_{2}[\bar A]=\frac{1}{2}{\rm Tr}\ln {\cal O}_{\mu\nu}^{\rm glue}-{\rm Tr}\ln {\cal O}^{\rm ghost}- \frac{N_f}{2} {\rm Tr}\ln \slashed{D}^2\,,
  \ee
  where the traces collectively denote traces are over Lorentz, color, spinor and space--time indices where appropriate. Using notation ${\cal A}_\mu^{ab}=f^{acb}\bar A_\mu^c$, ${\cal F}_{\mu\nu}^{ab}=f^{acb} \bar F_{\mu\nu}^c$, the operators ${\cal O}_{\mu\nu}^{\rm glue},{\cal O}^{\rm ghost}$ in momentum space are given by
  \begin{eqnarray}
     {\cal O}^{\rm ghost}&=&p^2 \delta^{ab}-2 i p_\mu {\cal A}_\mu^{ab}-\left({\cal A}^{2}\right)^{ab}\,,\quad\\
     {\cal O}_{\mu\nu}^{\rm glue}&=&\left(p^2 \delta^{ab}-2 i p_\alpha {\cal A}_\alpha^{ab}-\left({\cal A}^{2}\right)^{ab} \right) \delta_{\mu\nu}-2 {\cal F}_{\mu\nu}^{ab}\,.
     \nonumber
  \end{eqnarray}

  While it would clearly be desirable, it is not known how to evaluate the effective one-loop action for arbitrary space-time dependent background gauge field configurations. For this reason, we resort to a second (uncontrolled) approximation in our calculation by restricting the background field-strength tensor $\bar F_{\mu\nu}^a$, to be constant. This will allow us to calculate the logarithms of the operators using the heat-kernel method, cf. \cite{Blau:1988iz,Strassler:1992zr,Reuter:1996zm}. To simplify the calculation further, we will restrict ourselves to self-dual background field configurations, where the Euclidean chromo-electric and chromo-magnetic fields are taken to be equal, $\bar E^c=\bar B^c$. While this restriction is not necessary for the calculation itself, it is key in avoiding infrared instabilities such as those associated with the Nielsen-Olesen instability \cite{Nielsen:1978rm}.
  
  For positive-definite operators ${\cal O}$, we write
  \be
  {\rm Tr}\ln {\cal O}=-\int_0^\infty \frac{ds}{s} K(s)\,, \quad K(s)={\rm Tr} e^{-s {\cal O}}\,,
  \ee
  where the kernel $K(s)$ can further be written as a quantum-mechanical partition function
  \be
  K(s)={\rm tr}\int dx \left<x|e^{-s {\cal O}}|x\right>={\rm tr}\int {\cal D}x e^{-\Sigma}\,,
  \ee
  where the remaining trace has to be taken over color, spinor and Lorentz indices where appropriate. We find
  \begin{eqnarray}
    \label{sigmas}
    \Sigma^{\rm ghost}&=&\int_0^{s} d\sigma \left[\frac{\dot x_\mu^2}{4}\delta^{ab}+ \dot{x}_\mu {\cal A}_\mu^{ab}\right]\,,\\
    \Sigma^{\rm glue}&=&\int_0^{s} d\sigma \left[\frac{\dot x_\alpha^2}{4}\delta^{ab}\delta_{\mu\nu}+ \dot{x}_\alpha {\cal A}_\alpha^{ab} \delta_{\mu\nu}-2 {\cal F}_{\mu\nu}^{ab}\right]\,,\nonumber\\
    \Sigma^{\rm fermion}&=&\int_0^{s} d\sigma \left[\frac{\dot x_\alpha^2}{4}+ i \dot{x}_\alpha \bar A_\alpha^{a} t^a -\frac{i}{2}\sigma_{\mu\nu} \bar F_{\mu\nu}^{a} t^a\right]\,,\nonumber
  \end{eqnarray}
  where we have used $\slashed{D}^2=D^2-\frac{i}{2}\sigma_{\mu\nu}\bar F_{\mu\nu}^a t^a$ with $\sigma_{\mu\nu}=\frac{1}{2}\left[\gamma_\mu,\gamma_\nu\right]$ for the fermion operator, and the boundary conditions are $x_\mu(\sigma=0)=x_\mu(\sigma=s)$. We can write $\bar A_\mu^a(x)=-\frac{1}{2}\bar F_{\mu\nu}^a x_\nu$, so that $\Sigma$ is quadratic in $x_\mu$, and the path integral over $x$ can be done in all cases using methods from finite-temperature field theory \cite{Laine:2016hma}. Specifically, for the ghost kernel, we have for self-dual field configurations
  \be
  K_{\rm ghost}(s)={\cal C}\times {\rm tr} \prod_{n}\frac{1}{\omega_n^2 \delta^{ab}-4 f^{acd} f^{deb}\bar B^c \bar B^e}\,,
  \ee
  where $\omega_n=\frac{2 \pi n}{s}$ and we can fix the overall (divergent) constant ${\cal C}$ from calculating
\be
  K^{\bar A=0}_{\rm ghost}(s)={\rm tr}\int \frac{d^d p}{(2\pi)^d}e^{-s p^2 \delta^{ab}}={\rm vol}\times \frac{N^2-1}{(4 \pi s)^{\frac{d}{2}}}\,,
  \ee
  where ${\rm vol}$ denotes the space-time volume. We thus obtain
  \be
  K_{\rm ghost}(s)=\frac{\rm vol}{(4 \pi s)^{\frac{d}{2}}}{\rm tr}\frac{s^2{\cal M}}{4\sinh^2\left(\frac{s \sqrt{\cal M}}{2}\right)}\,,
  \ee
  where ${\cal M}=-4 f^{acd} f^{deb}\bar B^c \bar B^e$. The color trace over the matrix-valued function can be evaluated through the formula
  \be
  {\rm tr}\left({\cal M}^n\right)=\sum_{i=1}^8 \lambda_i^{n}\,,
  \ee
  where $\lambda_i$ are the eigenvalues of ${\cal M}$. For SU(3), we find that the eigenvalues come in pairs, and that two of the eigenvalues vanish $\lambda_0=0$, such that the only non-vanishing eigenvalues can be taken as $\lambda_1,\lambda_2,\lambda_3$. We thus have
  \be
  K_{\rm ghost}(s)=\frac{\rm vol}{(4 \pi s)^{\frac{d}{2}}} \sum_{i=0}^3\frac{s^2\lambda_i}{2\sinh^2\left(\frac{s \sqrt{\lambda_i}}{2}\right)}\,.
  \ee

For the gluons, since $\bar F_{\mu\nu}$ is constant, there is an additional factor of $e^{2 s {\cal F}_{\mu\nu}^{ab}}$ and an additional Lorentz trace in the kernel. Since the other parts of the kernel are diagonal in Lorentz indices, as well as diagonal in color, we can proceed to take the color trace, and obtain for the gluon kernel
\be
K_{\rm glue}(s)= \frac{{\rm vol}\times d}{(4 \pi s)^{\frac{d}{2}}}\sum_{i=0}^3 \frac{s^2 \lambda_i \cosh\left(s \sqrt{\lambda_i}\right)}{2\sinh^2\left(\frac{s \sqrt{\lambda_i}}{2}\right)}
  \ee

  Finally, for the fermions, expanding the exponent involving $\sigma\cdot F$ in a power series, only even powers survive the spinor trace. Using furthermore $\left(\sigma\cdot F\right)^2=F_{\mu\nu}F_{\alpha\beta}\frac{1}{2}\left\{\sigma_{\mu\nu},\sigma_{\alpha \beta}\right\}$ and (cf. \cite{Hattori:2020guh})
  \be
  \left\{\sigma_{\mu\nu},\sigma_{\alpha\beta}\right\}=2\left(\delta^{\mu\beta}\delta^{\nu\alpha}-\delta^{\mu\alpha}\delta^{\nu\beta}+\epsilon^{\mu\nu\alpha\beta}\gamma_5\right)\,,
  \ee
  for Euclidean gamma matrices, we get  for self-dual field configurations 
  \be
  K_{\rm fermion}(s)= \frac{{\rm vol}\times d}{(4 \pi s)^{\frac{d}{2}}}\sum_{i=1}^3 \frac{s^2 \lambda_{F,i} \cosh^2\left(\frac{s \sqrt{\lambda_{F,i}}}{2}\right)}{4\sinh^2\left(\frac{s \sqrt{\lambda_{F,i}}}{2}\right)}\,,
  \ee
  where $\lambda_{F,i}$ are the eigenvalues of ${\cal M}_F=4 t^a t^b \bar B^a \bar B^b$.

  Putting everything together, we have $K_{\rm sum}(s)=\frac{1}{2}K_{\rm glue}(s)-K_{\rm ghost}(s)-\frac{N_f}{2} K_{\rm fermion}(s)$, and we note that $K_{\rm sum}$ contains the terms
  \be
  \label{ssss}
  \frac{s^2d \times {\rm vol}}{2 (4 \pi s)^\frac{d}{2}}\sum_{i=1}^3 \left(\lambda_i-\frac{N_f}{4}\lambda_{F,i}\right)\,,
  \ee
  that lead to an infrared divergence for general values of $N,N_f$. However, noting that sums over eigenvalues are related to the color traces over the adjoint and fundamental representation matrices,
  \be
  \label{identity}
  \sum_{i=1}^3 \lambda_i=\frac{1}{2}{\rm tr} {\cal M}=2 N \bar B^a \bar B^a\,,\quad
  \sum_{i=1}^3 \lambda_{F,i}={\rm tr} {\cal M}_F=2 \bar B^a \bar B^a
  \ee
so that we find that for the special choice of
  \be
  \label{choice}
  N_f=4 N\,,
  \ee
the infrared divergence exactly cancels between the gluonic and fermionic degrees of freedom, irrespective of the choice for the non-abelian background field $\bar B^a$. Therefore, we choose to focus on QCD with twelve massless fermion flavors in the following.

We get for the full kernel 
\begin{eqnarray}
  \label{ksum}
  K_{\rm sum}&=&\frac{\rm vol}{(4\pi s)^{\frac{d}{2}}}\left[(d-2)\right.\\
    &+&\left.\sum_{i=1}^3
\left(\frac{(d-2)s^2 \lambda_i}{4\sinh^2\left(\frac{s \sqrt{\lambda_i}}{2}\right)}
  -\frac{N_f d}{2}\frac{s^2\lambda_{F,i}}{4 \sinh^2\left(\frac{s \sqrt{\lambda_{F,i}}}{2}\right)}\right)\right]\,,\nonumber
\end{eqnarray}
where the term in the first line does not to contribute to the effective action in dimensional regularization except in a medium. Using \cite[25.5.9]{NIST:DLMF}, we have
  \be
  \int_0^\infty \frac{ds}{(4 \pi s)^{\frac{d}{2}}} \frac{s}{\sinh^2(s)}=\left(\frac{1}{2\pi}\right)^{\frac{d}{2}}\zeta\left(1-\frac{d}{2}\right)\Gamma\left(2-\frac{d}{2}\right)\,,
  \ee
  and we therefore obtain for the effective action of twelve flavor QCD in dimensional regularization
  \begin{eqnarray}
    \frac{S_2[\bar A]}{\rm vol}&=&\frac{1}{(4\pi)^2}\left[\frac{-N \bar B^a \bar B^a}{\varepsilon}+12 N \bar B^a \bar B^a\zeta^\prime(-1)\right.\\
      &&\left.
      +\sum_{i=1}^3\left(\frac{\lambda_i\ln \frac{\bar \mu^2e^{-1}}{\sqrt{\lambda_i}}-N_f \lambda_{F,i}\ln \frac{\bar\mu^2 e^{-\frac{1}{2}}}{\sqrt{\lambda_{F,i}}}}{6}\right)
      \right]\,,\nonumber
  \end{eqnarray}
  where we used again (\ref{identity}) and where $\bar\mu^2=4 \pi \mu^2 e^{-\gamma_E}$ is the $\overline{\rm MS}$ parameter. Using $S_0={\rm vol}\times \frac{\bar B^a \bar B^a}{g_0^2}$, we can remove all divergences by non-perturbatively renormalizing the coupling as
  \be
  \frac{1}{g_0^2}=\frac{1}{g_R^2(\bar\mu)}+\frac{N}{(4\pi)^2 \varepsilon}\,,
  \ee
  so that the running coupling becomes
  \be
  \label{exrun}
  \frac{1}{g_R^2(\bar\mu)}=\frac{N}{(4\pi)^2}\ln \frac{\bar\mu^2}{\Lambda_{\overline{\rm MS}}^2}\,.
  \ee
  Note that the associated $\beta$ function matches the one-loop result $\beta_0=\frac{11 N}{3}-\frac{2 N_f}{3}$ when letting $N_f=4 N$, cf. \cite{vanRitbergen:1997va}. For a physical interpretation of (\ref{exrun}) in the infrared, cf. Refs.~\cite{Symanzik:1973hx,Romatschke:2022jqg,Romatschke:2022llf,Romatschke:2023sce,Romatschke:2023fax,Weller:2023jhc}. Using the explicit from of the running coupling (\ref{exrun}), the renormalized partition function thus becomes $Z=\int d\bar B e^{-V_{\rm eff}[\bar B]}$ with
  \begin{eqnarray}
    \label{veff}
    \frac{V_{\rm eff}[\bar B]}{\rm vol}&=&\frac{1}{(4\pi)^2}\left[12 N \bar B^a \bar B^a \zeta^\prime(-1)-\frac{N \bar B^a \bar B^a}{6}\right.\\
    -&&\left.\sum_i\left(\frac{\lambda_i}{12}\ln \left(\frac{\lambda_i e^{1}}{\Lambda^4_{\overline{\rm MS}}}\right)-\frac{N_f \lambda_{F,i}}{12}\ln \left(\frac{\lambda_{F,i}e^{1}}{\Lambda^4_{\overline{\rm MS}}}\right)\right)\right]\,.
\nonumber  \end{eqnarray}
  This result is independent of the fictitious renormalization scale $\bar\mu$, as it should be for a physical observable.

  In the large volume limit, we can solve $Z$ using the method of steepest descent. It is easy to identify the trivial saddle point as $\bar B^a=0$, with $\ln Z_{\rm triv}=0$. This saddle can be recognized to be the starting point of perturbative QCD calculations.
 
Besides the trivial saddle, (\ref{veff}) also contains non-trivial saddles. This is most easily seen by considering the abelian sub-sectors of SU(3), with $\bar B^a$ directed along a single direction $a=1,2,\ldots,8$. We find that there are two inequivalent abelian subsectors, those with $a=1,2,\ldots 7$, and one with $a=8$ if using Gell-Mann matrices for the generators. The corresponding saddle locations are
\begin{eqnarray}
    \delta^{a1}:\, \bar B&=&\sigma_1=e^{-12 \zeta^\prime(-1)-\frac{5}{6}}{2^{\frac{2}{9}}}\Lambda^2_{\overline{\rm MS}}\simeq 3.69 \Lambda^2_{\overline{\rm MS}}\,,\\
    \delta^{a8}:\, \bar B&=&\sigma_8=e^{-12 \zeta^\prime(-1)-\frac{5}{6}}{2^{-\frac{8}{9}}3^{\frac{5}{6}}}\Lambda^2_{\overline{\rm MS}}\simeq 4.27 \Lambda^2_{\overline{\rm MS}}\,,\nonumber
  \end{eqnarray}
with free energy $-\frac{\ln Z_{1,8}}{\rm vol}=-\frac{3 \sigma_{1,8}^2}{32 \pi^2}$, respectively. Since $-\frac{\ln Z}{\rm vol}$ is the free energy density of the theory, and since saddles with lower free energy dominate, it is clear that the abelian subsector along the color direction $a=8$ dominates over both the other abelian subsectors, and also over the trivial perturbative saddle with $\bar B=0$. For non-abelian configuration $\bar B^a$, there are additional saddles, and we have not succeeded in a full classification of all of these saddles. However, for the special case of $\bar B^a=\bar B_1 \delta^{a1}+\bar B_8 \delta^{a8}$ (linear combination of two color directions), we find a nontrivial saddle for $\bar B_1=\frac{\sqrt{3}\sigma_8}{2},\bar B_8=\frac{\sigma_8}{2}$, with the result  $\frac{\ln Z_{mix}}{\rm vol}=\frac{3 \sigma_{8}^2}{32 \pi^2}$. We again encounter the same value for the free energy when finding the saddle for a linear combination of three, four and five color directions, e.g. $\bar B^a=\sum_{b=1}^4 \bar B_b \delta^{ab}+\bar B_8 \delta^{a8}$. For this reason, we believe that the non-trivial saddle in the abelian subsector with $\bar B^a=\delta^{a8}\sigma_8$ either is, or is close to, the true saddle for the non-abelian effective potential at zero temperature.

\subsection{Finite Temperature}

  At finite temperature $T$, the Euclidean time direction is compactified on a circle with circumference $\beta\equiv \frac{1}{T}$. In the present approach, this implies non-trivial boundary conditions for the kernel $K(s)$, cf. Refs.~\cite{McKeon:1992if,Shovkovy:1998xw}. Specifically, for (\ref{sigmas}) we have
  \be
  \label{notriva}
  x_0(s)=x_0(0)+n \beta,\qquad n\in \mathbb{N}\,,
  \ee
  at finite temperature. The path integral can be calculated using the Gelfand-Yaglom method, cf. Ref.~\cite{Dunne:2007rt}, with the finite temperature contribution arising entirely from the classical contribution to (\ref{sigmas}).  Since the spatial components obey periodic boundary conditions $x_i(0)=x_i(s)$, the classical equations of motion imply $x_i=0$, whereas the only non-trivial field must fulfill $\ddot x_0=0$. Together with the boundary conditions (\ref{notriva}), one finds that the classical action for ghosts, gluons and fermions each is given by
  \be
  \Sigma_{\rm cl}=\left.\frac{x_0 \dot x_0}{4}\right|_{0}^{s}=\frac{n^2 \beta^2}{4 s}\,.
  \ee
  As a consequence, the finite-temperature kernels for ghosts, gluons and fermions are given by
  \be
  \label{finiteT}
  K_{\pm}(s)=\sum_{n=-\infty}^\infty K_{\pm}^{T=0}(s) (\pm 1)^{n} e^{-\frac{n^2 \beta^2}{4 s}}\,,
  \ee
  with plus for ghosts and gluons and minus for fermions, and the $T=0$ superscript denotes the zero-temperature expressions from above.

One can immediately recognize the zero-temperature result from before to correspond to the term $n=0$ in the infinite series in (\ref{finiteT}). In addition, one finds that for all the odd terms in the series, the finite temperature contribution for the fermions in $K_{\rm sum}(s)$ is the same sign as the gluon contribution. This implies that (\ref{ssss}) at finite temperature possesses an uncancelled infrared divergence. This was to be expected: at finite temperature, fermions and gluons receive different in-medium mass corrections, and therefore the delicate cancellation obtained for (\ref{choice}) no longer works at finite temperature, as was already pointed out in the introduction.

We do not have calculational control over the IR - divergent terms, but we do have control over the other finite temperature contributions. An uncontrolled approximation consists of simply dropping the IR divergent terms and only keeping the finite terms of the effective potential. We do this to try to get at least \textit{some} estimate of finite-temperature effects by restricting our calculation to the sector where (\ref{choice}) removes all infrared divergencies: the even values of $n$ in (\ref{finiteT}). While not suitable for calculating quantitative observables in this theory, restricting to this thermal subsector does allow us to perform a consistent finite-temperature calculation not affected by infrared divergencies. With this restriction, the finite-temperature generalization of (\ref{ksum}) becomes
\begin{eqnarray}
  \label{Tksum}
  K_{\rm sum}^{\rm rest.}&=&\frac{\rm vol}{(4\pi s)^{\frac{d}{2}}}\sum_{n=-\infty}^\infty e^{-\frac{n^2 \beta^2}{s}}\left[(d-2)\right.\\
    \left.+\sum_{i=1}^3\right.&&\left.
\left(\frac{(d-2)s^2 \lambda_i}{4\sinh^2\left(\frac{s \sqrt{\lambda_i}}{2}\right)}
-\frac{N_f d }{2}\frac{s^2\lambda_{F,i}}{4 \sinh^2\left(\frac{s \sqrt{\lambda_{F,i}}}{2}\right)}\right)\right]\,.\nonumber
%&&+\frac{4 s^2 N d \bar B^2\times {\rm vol}}{(4 \pi s)^\frac{d}{2}} \sum_{n=0}^\infty e^{-\frac{(2n+1)^2 \beta^2}{4 s}-m^2 s}\,.\nonumber
\end{eqnarray}

Using
\be
\int_0^\infty \frac{ds}{s} \frac{e^{-\frac{n^2 \beta^2}{s}}}{ \sinh^2(\frac{\sqrt{\lambda} s}{2})} =8 \sum_{l=1}^\infty l K_0\left(2 n \beta \lambda^{\frac{1}{4}}\sqrt{l}\right)\,,
\ee
with $K_0(x)$ denoting a modified Bessel function of the second kind,  we can evaluate all the integrals over $s$. Renormalizing the UV divergencies as before, we find for the restricted finite-temperature effective potential $\Delta V^{\rm rest}_{\rm eff}=V^{\rm rest}_{\rm eff}[\bar B,T]-V_{\rm eff}[\bar B,0]$ the result
\begin{eqnarray}
    \label{Tveff}
    \frac{\Delta V_{\rm eff}^{\rm rest}[\bar B,T]}{\rm vol}&=&-\sum_{n=1}^\infty\sum_{i=1}^3\sum_{l=1}^\infty \frac{l A(n,i,l)}{2\pi^2}\\
    A(n,i,l)=&&\!\!\!\!\!\!\!\lambda_i K_0\left(\frac{2 n \sqrt{l} \lambda_i^{\frac{1}{4}}}{T}\right)-N_f   \lambda_{F,i} K_0\left(\frac{2 n \sqrt{l} \lambda_{F,i}^{\frac{1}{4}}}{T}\right)\,,\nonumber
\end{eqnarray}
where we dropped a $\bar B$-independent contribution.
We can numerically search for minima of the finite-temperature effective potential. For the abelian subsector $\bar B^a=\bar B \delta^{a8}$, we find that raising the temperature initially results in a minimum for the effective potential that is close to $\Bar B=\sigma_8$ until a critical temperature of
\be
T_c\simeq 0.81 \Lambda_{\overline{\rm MS}}\,,
\ee
above which the stable minimum jumps to $\bar B=0$. Since the location jumps discontinuously at $T=T_c$, this indicates a first order phase transition for thermodynamic properties.

  \section{Results and Discussion}

  Since the non-perturbative non-abelian saddles have lower free energy than the perturbative saddle, in the large volume limit we find
  \be
  \label{Z8}
  \frac{\ln Z}{\rm vol}=\frac{3\sigma_8^2}{32\pi^2}\,.
  \ee
  Comparing the free energy for the perturbative saddle $\bar B=0$ to (\ref{Z8}), we obtain the non-perturbative value of the bag constant:
  \be
  C_{\rm bag}=-\frac{3\sigma_8^2}{32\pi^2}\simeq -0.173 \Lambda^4_{\overline{\rm MS}}\,.
  \label{cbag}
 \ee
 Note that in the original phenomenological bag model of hadrons \cite{Chodos:1974je}, the bag constant is taken to be positive, whereas in our calculation it comes out negative.  In fact, it seems that thermodynamic stability requires the sign of the bag constant (\ref{cbag}) to be negative. This is because a positive sign for $C_{\rm bag}$ would imply that the non-perturbative saddle $\bar B\neq 0$ is thermodynamically disfavored with respect to the perturbative vacuum at $\bar B=0$.

 We can also evaluate the Polyakov loop of length L
 \be
 \label{polyakov}
 P(L,\vec{x})= {\cal P} e^{i \oint dx A_0^a(\vec{x}) t^a}\,,
 \ee
 for our saddle-point solution $A_\mu^a=\bar A_\mu^a=-\frac{1}{2} \bar F_{\mu\nu}^a x_\nu$ where ${\cal P}$ denotes path ordering. For the abelian subsector $\bar B^a=\sigma_8 \delta^{a8}$, we find
 \be
 \frac{1}{N}{\rm tr} P(L,\vec{x})=\frac{\cos\left(\frac{\sigma_8 L (x+y+z)}{\sqrt{3}}\right)}{3}+\frac{2\cos\left(\frac{\sigma_8 L (x+y+z)}{\sqrt{12}}\right)}{3}\,,
 \ee
 so that
 \be
 \int d^3\vec{x} \frac{1}{N}{\rm tr} P(L,\vec{x})=0\,.
 \ee
 Note that we recover identical results for the Polyakov loop for other non-abelian saddles $\bar B^a$ as long as they have full rank. Our restricted finite temperature evaluation of the theory indicates that the saddle jumps discontinuously to $\bar B=0$ for $T>T_c$, so that $\frac{1}{N}{\rm tr} P(L,\vec{x})=1$ in the high temperature region. The behavior found in our calculation is consistent with that of a theory that exhibits color confinement at low temperatures and a deconfinement transition taking place at $T=T_c$  \cite{Lenz:2003jp,Fodor:2017gtj}. We caution that a vanishing Polyakov loop vacuum expectation value is a necessary, but not a sufficient condition for color confinement.

 While our results are consistent with confinement occurring in $N_f=12$ flavor QCD, other studies using lattice QCD and perturbation theory have argued that the theory is instead infrared conformal\cite{DeGrand:2015zxa,DiPietro:2020jne}. The most recent works on this subject combine both unambiguous lattice QCD results with at $N_f$ with perturbative studies at high $N_f$ and find that QCD with $N_f=12$ flavors is ``around where the conformal window starts'' \cite{Chung:2023mgr,Chung:2023iwn}.

The presence of additional saddles strongly suggests a more complicated phase structure of the theory. For instance, for the saddle along the abelian subsector $\bar B^a=\sigma_3 \delta^{a3}$, the corresponding generator $t^3$ in (\ref{polyakov}) does not have full rank. As a consequence, the expectation value for $\int d^3\vec{x} \frac{1}{N}{\rm tr} P(L,\vec{x})=\frac{1}{3}$, suggesting that for this saddle, the theory is neither fully confined, nor fully deconfined. We suspect that this saddle corresponds to the partially deconfined phase discussed in Refs.~\cite{Hanada:2018zxn,Hanada:2022wcq,Hanada:2023krw}.

To summarize, we have presented a non-perturbative calculation of the one-loop effective potential for QCD with twelve flavors of massless fermions that is free of ultraviolet and infrared divergencies. We found that in addition to the perturbative vacuum, the effective potential possesses  non-perturbative saddles with lower free energy. We calculated the expectation value of the Polyakov loop for different saddles, finding that it vanishes identically for the non-perturbative saddle with the lowest free energy. We also found explicit non-perturbative saddles with expectation values for the Polyakov loop of $\frac{1}{3}$, possibly corresponding to a partially deconfined phase. In addition, we presented a restricted analysis of finite-temperature effects in the theory, within which we found a first-order transition to the perturbative vacuum at a calculable critical temperature $T_c$. Many extensions of our work are feasible, such as calculations at finite density, or calculations of real-time quantities. We leave such endeavors for future work.

\section*{Acknowledgments}

We would like to thank Scott Lawrence for encouraging to write up this result and Daniel Nogradi for helpful background information on $N_f=12$ lattice QCD. This work was supported by the Department of Energy, DOE award No DE-SC0017905. 

\bibliography{lit}
\end{document}